\pdfoutput=1

\documentclass[11pt,twoside,a4paper,cmspaper,final,collab]{cms-tdr}
\def\svnVersion{339282}\def\cmsCernNoTag{CERN-PH-EP/2015-192}\def\cmsCernDate{\today}\def\cmsMessage{Published in Physics Letters B as \href{http://dx.doi.org/10.1016/j.physletb.2016.03.046}{\doi{10.1016/j.physletb.2016.03.046.}}}

\begin{document}\cmsNoteHeader{BPH-13-012}

\hyphenation{had-ron-i-za-tion}
\hyphenation{cal-or-i-me-ter}
\hyphenation{de-vices}

\RCS$Revision: 339229 $
\RCS$HeadURL: svn+ssh://svn.cern.ch/reps/tdr2/papers/BPH-13-012/trunk/BPH-13-012.tex $
\RCS$Id: BPH-13-012.tex 339229 2016-04-19 10:59:35Z gfedi $

\newlength\cmsFigWidth
\ifthenelse{\boolean{cms@external}}{\setlength\cmsFigWidth{0.48\textwidth}}{\setlength\cmsFigWidth{0.6\textwidth}}
\ifthenelse{\boolean{cms@external}}{\providecommand{\cmsLeft}{top\xspace}}{\providecommand{\cmsLeft}{left\xspace}}
\ifthenelse{\boolean{cms@external}}{\providecommand{\cmsRight}{bottom\xspace}}{\providecommand{\cmsRight}{right\xspace}}

\newcommand{\BsJpsiPhi}{\ensuremath{\mathrm{B^0_s} \to \JPsi\,\phi(1020)}\xspace}
\newcommand{\BsJpsiFz}{\ensuremath{\mathrm{B^0_s} \to  \JPsi\,\mathrm{f_{0}}(980)}\xspace}

\newcommand{\Bu}{\ensuremath{\mathrm{B}^\pm}\xspace}
\newcommand{\Bs}{\ensuremath{\mathrm{B}^0_s}\xspace}
\newcommand{\Bd}{\ensuremath{\mathrm{B}^0}\xspace}
\newcommand{\Bc}{\ensuremath{\mathrm{B_c}}\xspace}
\newcommand{\Bsbar}{\ensuremath{\mathrm{\overline{B}{}^0_s}}\xspace}
\newcommand{\Lb}{\ensuremath{\Lambda_\mathrm{b}}\xspace}
\newcommand{\JpsiPhi}{\ensuremath{\JPsi\phi}\xspace}
\newcommand{\deltagammas}{\ensuremath{\Delta \Gamma_\mathrm{s}}\xspace}
\providecommand{\NA}{\ensuremath{\text{---}\xspace}}

\title{Measurement of the CP-violating weak phase $\phi_\mathrm{s}$ and the decay width difference $\Delta\Gamma_\mathrm{s}$ using the ${\mathrm{B^0_s} \to \JPsi\,\phi(1020)}$ decay channel in pp collisions at $\sqrt{s}= 8\TeV$}

\date{\today}

\abstract{
The CP-violating weak phase $\phi_\mathrm{s}$ of the \Bs{} meson and the decay width difference $\Delta\Gamma_\mathrm{s}$ of the \Bs{} light and heavy mass eigenstates are measured with the CMS detector at the LHC using a data sample of ${\mathrm{B^0_s} \to  \JPsi\,\phi(1020)} \to \mu^+\mu^-\PKp\PKm$ decays. The analysed data set corresponds to an integrated luminosity of 19.7\fbinv collected in pp collisions at a centre-of-mass energy of 8\TeV.
A total of 49\,200 reconstructed \Bs{} decays are used to extract the values of $\phi_\mathrm{s}$ and $\Delta\Gamma_\mathrm{s}$ by performing a time-dependent and flavour-tagged angular analysis of the $\mu^+ \mu^- \PKp\PKm$ final state. The weak phase is measured to be  $\phi_\mathrm{s} = -0.075 \pm 0.097\stat\pm 0.031\syst\unit{rad}$, and the decay width difference is $\Delta\Gamma_\mathrm{s} = 0.095 \pm 0.013\stat\pm 0.007\syst\unit{ps}^{-1}$.
}

\hypersetup{%
pdfauthor={CMS Collaboration},%
pdftitle={Measurement of the CP-violating weak phase phi[s] and the decay width difference DeltaGamma[s] using the Bs to J/Psi phi(1020) decay channel in pp collisions at sqrt(s) = 8 TeV},%
pdfsubject={CMS},%
pdfkeywords={CMS, physics, B-physics, CP violation}}

\maketitle

\section{Introduction}
While no direct evidence of physics beyond the standard model (SM) has yet been found at the CERN LHC, the \Bs{} meson provides a rich source of possibilities to probe its consistency.
In this Letter, a measurement of the weak phase $\phi_\mathrm{s}$ of the \Bs{} meson and the decay width difference $\Delta\Gamma_\mathrm{s}$ between the light and heavy \Bs{} mass eigenstates is presented, using the data collected by the CMS experiment in pp collisions at the LHC with a centre-of-mass energy of $8\TeV$, corresponding to an integrated luminosity of 19.7\fbinv.

The CP-violating weak phase $\phi_\mathrm{s}$ originates from the interference between direct \Bs{} meson decays into a CP eigenstate $\PQc\PAQc\PQs\PAQs$ and decays through \Bs{}--\Bsbar{} mixing to the same final state.
Neglecting penguin diagram contributions~\cite{S0217751X13500632}\cite{Faller:2008gt}, $\phi_\mathrm{s}$ is related to the elements of the Cabibbo--Kobayashi--Maskawa quark mixing matrix by $\phi_\mathrm{s} \simeq -2\beta_\mathrm{s}$, where $\beta_\mathrm{s}=\arg(-V{}_\mathrm{\!ts}V^*{}_\mathrm{\!\!\!tb}/V{}_\mathrm{\!cs}V^*{}_\mathrm{\!\!\!cb})$. The prediction for $2\beta_\mathrm{s}$, determined via a global fit to experimental data within the SM, is $2\beta_\mathrm{s}=0.0363\,^{+0.0016}_{-0.0015}\unit{rad}$~\cite{Charles:2011va}. Since the value predicted by the SM is very precise, any significant deviation of the measured value from this prediction would be particularly interesting, as it would indicate a possible contribution of new, unknown particles to the loop diagrams describing \Bs{}  mixing.
The theoretical prediction for the decay width difference $\Delta\Gamma_{\mathrm{s}}$ between the light and heavy \Bs{} mass eigenstates $\mathrm{B}_\mathrm{L}$ and $\mathrm{B}_\mathrm{H}$, assuming no new physics in \Bs{}--\Bsbar{} mixing, is  $\Delta\Gamma_{\mathrm{s}} = \Gamma_{\mathrm{L}}- \Gamma_{\mathrm{H}}= 0.087\pm 0.021\unit{ps}^{-1}$~\cite{Lenz:2011ti}.

The weak phase $\phi_\mathrm{s}$ was first measured by the Tevatron experiments~\cite{Aaltonen:2007he, Abazov:2008af, CDF:2011af, Abazov:2011ry}, and then at the LHC by the LHCb and ATLAS experiments~\cite{Aaij:2014oba, Aaij:2013oba, LHCb:2012ad, Aad:2012kba,Aad:2014cqa}, using \BsJpsiPhi{}, ${\mathrm{B^0_s} \to  \JPsi\,\mathrm{f_{0}}(980)}$, and ${\mathrm{B^0_s}\to \JPsi\,\pi^+ \Pgpm}$ decays to $\ell^+ \ell^- \mathrm{h}^+ \mathrm{h}^-$, where $\ell$ denotes a muon in the present analysis and h stands for a kaon or a pion. Final states that do not have a single CP eigenvalue require an angular analysis to disentangle the CP-odd and CP-even components. The time-dependent angular analysis can be performed by measuring the decay angles of the final-state particles $\ell^+ \ell^- \mathrm{h}^+\mathrm{h}^-$ and the proper decay time of the \Bs{} multiplied by the speed of light~\cite{Dighe:1999}, referred to as $ct$ in what follows.
In this Letter, the \BsJpsiPhi{} decay to the final state $\mu^+\mu^-\, \PKp\PKm$ is analysed, and possible additional contributions to the result from the nonresonant decay $\mathrm{B^0_s}\to \JPsi\, \PKp\PKm$ are taken into account by including a term for an additional amplitude ($S$-wave) in the fit.

\begin{figure}[hbtp]
  \centering
    \includegraphics[width=0.48\textwidth]{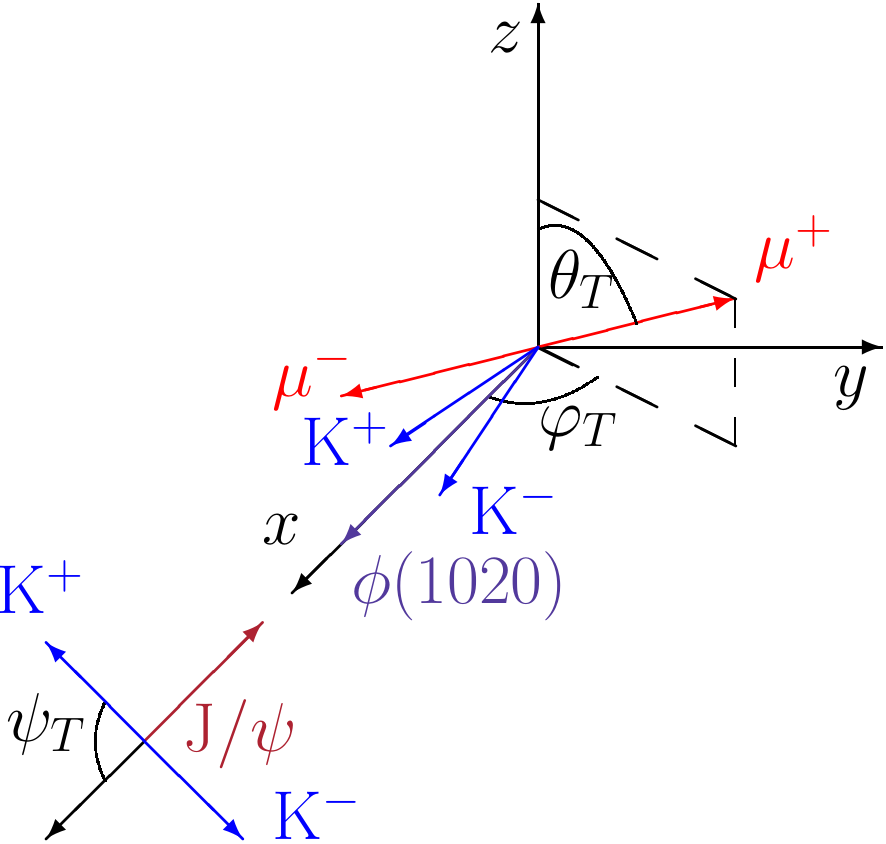}
    \caption{Definition of the three angles $\theta_\mathrm{T}$, $\psi_\mathrm{T}$, and $\varphi_\mathrm{T}$ describing the decay topology of \BsJpsiPhi. See text for details.}
    \label{fig:ex1}
\end{figure}

In this measurement the transversity basis is used~\cite{Dighe:1999}. The three angles $\Theta=(\theta_\mathrm{T},\psi_\mathrm{T},\varphi_\mathrm{T})$ of the transversity basis are illustrated in Fig.~\ref{fig:ex1}. The angles $\theta_\mathrm{T}$ and $\varphi_\mathrm{T}$ are the polar and azimuthal angles, respectively, of the $\mu^+$ in the rest frame of the \JPsi{} where the $x$ axis is defined by the direction of the $\phi$(1020) meson in the \JPsi{} rest frame, and the $x$-$y$ plane is defined by the decay plane of the $\phi(1020)\to \PKp\PKm$. The helicity angle $\psi_\mathrm{T}$ is the angle of the $\mathrm{K^+}$ in the $\phi(1020)$ rest frame with respect to the negative \JPsi{} momentum direction.

The differential decay rate of \BsJpsiPhi{} is represented using the function $f(\Theta,ct,\alpha)$ as in  Ref.~\cite{Dighe:1995pd}:
\begin{equation}
\frac{\rd^4\Gamma\!\left(\mathrm{B^0_s}\right)}{\rd\Theta\,\rd\!\left(ct\right)} = f(\Theta,ct,\alpha)\propto \sum^{10}_{i=1}O_i(ct,\alpha)\, g_i(\Theta),
\label{eqnarray:decayrate}
\end {equation}

where $O_i$ are time-dependent functions, $g_i$ are angular functions, and $\alpha$ is a set of physics parameters.

The functions $O_i(ct, \alpha)$ are:
\ifthenelse{\boolean{cms@external}}{
\begin{multline*}
 O_i(ct,\alpha)=N_i \re^{- ct/c\tau}\Biggl[a_i\cosh\left(\frac{\Delta\Gamma_\mathrm{s} t}{2}\right)+b_i\sinh\left(\frac{\Delta\Gamma_\mathrm{s} t}{2}\right)\\
 +c_i\cos\left(\Delta m_\mathrm{s} t\right)+d_i\sin\left(\Delta m_\mathrm{s} t\right)\Biggr],
\label{eqnarray:observables}
\end{multline*}
}{
\begin{equation*}
 O_i(ct, \alpha)=N_i \re^{- ct/c\tau}\left[a_i\cosh\left(\frac{\Delta\Gamma_\mathrm{s} t}{2}\right)+b_i\sinh\left(\frac{\Delta\Gamma_\mathrm{s} t}{2}\right)+c_i\cos\left(\Delta m_\mathrm{s} t\right)+d_i\sin\left(\Delta m_\mathrm{s} t\right)\right],
\label{eqnarray:observables}
\end{equation*}
}
where $\Delta m_\mathrm{s}$ is the mass difference between the heavy and light \Bs{} mass eigenstates, $c\tau$ is defined as the product of the lifetime and the speed of light, the function $g_i(\Theta)$ and the terms $N_i$, $a_i$, $b_i$, $c_i$, and $d_i$ are given in Table~\ref{table:kinematics}.

\begin {table*}[h!tb]
\centering
\topcaption{Angular and time-dependent terms of the signal model.}
\label{table:kinematics}
\renewcommand{\arraystretch}{1.15}
\resizebox{\textwidth}{!}{
  \begin{tabular}{@{\hskip 0pt} l |@{\hskip 4pt} c @{\hskip 4pt}|@{\hskip 4pt} c @{\hskip 4pt}|@{\hskip 4pt} c @{\hskip 4pt}|@{\hskip 4pt} c @{\hskip 4pt}|@{\hskip 4pt} c @{\hskip 4pt}|@{\hskip 4pt} c @{\hskip 0pt}}
    $i$ & $g_i(\theta_\mathrm{T},\psi_\mathrm{T},\varphi_\mathrm{T})$ & $N_i$ & $a_i$ & $b_i$ & $c_i$ & $d_i$ \\
\hline
\rule{0pt}{3ex}
    \hspace{-0.2cm} 1 & $2\cos^2\psi_\mathrm{T}(1-\sin^2\theta_\mathrm{T}\cos^2\varphi_\mathrm{T})$                            & $\abs{A_0(0)}^2$                                  & 1 & $D$ & $C$ & $-S$\\
    2 & $\sin^2\psi_\mathrm{T}(1-\sin^2\theta_\mathrm{T}\sin^2\varphi_\mathrm{T})$                                & $\abs{A_{\parallel}(0)}^2$                    & 1 & $D$ & $C$ & $-S$\\
    3 & $\sin^2\psi_\mathrm{T}\sin^2\theta_\mathrm{T}$                                                            & $\abs{A_{\perp}(0)}^2$                          & 1 & $-D$ & $C$ & $S$\\
    4 & $-\sin^2\psi_\mathrm{T}\sin2\theta_\mathrm{T}\sin\varphi_\mathrm{T}$                                           & $\abs{A_{\parallel}(0)}\abs{A_{\perp}(0)}$ & $C\sin(\delta_{\perp}-\delta_{\parallel})$ & $S\cos(\delta_{\perp}-\delta_{\parallel})$ & $\sin(\delta_{\perp}-\delta_{\parallel})$ & $D\cos(\delta_{\perp}-\delta_{\parallel})$\\
    5 & $\frac{1}{\sqrt{2}}\sin2\psi_\mathrm{T}\sin^2\theta_\mathrm{T}\sin2\varphi_\mathrm{T}$              & $\abs{A_{0}(0)}\abs{A_{\parallel}(0)}$        & $\cos(\delta_{\parallel}-\delta_{0})$ & $D\cos(\delta_{\parallel}-\delta_{0})$ & $C\cos(\delta_{\parallel}-\delta_{0})$ & $-S\cos(\delta_{\parallel}-\delta_{0})$\\
    6 & $\frac{1}{\sqrt{2}}\sin2\psi_\mathrm{T}\sin2\theta_\mathrm{T}\sin\varphi_\mathrm{T}$                  & $\abs{A_{0}(0)}\abs{A_{\perp}(0)}$              & $C\sin(\delta_{\perp}-\delta_{0})$ & $S\cos(\delta_{\perp}-\delta_{0})$ & $\sin(\delta_{\perp}-\delta_{0})$ & $D\cos(\delta_{\perp}-\delta_{0})$\\
    7 & $\frac{2}{3}(1-\sin^2\theta_\mathrm{T}\cos^2\varphi_\mathrm{T})$                                   & $\abs{A_S(0)}^2$                                   & 1 & $-D$ & $C$ & $S$\\
    8 & $\frac{1}{3}\sqrt{6}\sin\psi_\mathrm{T}\sin^2\theta_\mathrm{T}\sin2\varphi_\mathrm{T}$              & $\abs{A_S(0)}\abs{A_{\parallel}(0)}$        & $C\cos(\delta_{\parallel}-\delta_{S})$ & $S\sin(\delta_{\parallel}-\delta_{S})$ & $\cos(\delta_{\parallel}-\delta_{S})$ & $D\sin(\delta_{\parallel}-\delta_{S})$\\
    9 & $\frac{1}{3}\sqrt{6}\sin\psi_\mathrm{T}\sin2\theta_\mathrm{T}\cos\varphi_\mathrm{T}$                 & $\abs{A_{S}(0)}\abs{A_{\perp}(0)}$              & $\sin(\delta_{\perp}-\delta_{S})$ & $-D\sin(\delta_{\perp}-\delta_{S})$ & $C\sin(\delta_{\perp}-\delta_{S})$ & $S\sin(\delta_{\perp}-\delta_{S})$\\
    10 & $\frac{4}{3}\sqrt{3}\cos\psi_\mathrm{T}(1-\sin^2\theta_\mathrm{T}\cos^2\varphi_\mathrm{T})$ & $\abs{A_{S}(0)}\abs{A_{0}(0)}$                    & $C\cos(\delta_{0}-\delta_{S})$ & $S\sin(\delta_{0}-\delta_{S})$ & $\cos(\delta_{0}-\delta_{S})$ & $D\sin(\delta_{0}-\delta_{S})$\\
\end{tabular}
}
\end{table*}

The terms $C$, $S$, and $D$ are defined as:
\begin{align*}
C&=\frac{1-\abs{\lambda}^2}{1+\abs{\lambda}^2}, &  S&=-\frac{2\abs{\lambda}\sin\phi_\mathrm{s}}{1+\abs{\lambda}^2},  & D&=-\frac{2\abs{\lambda}\cos\phi_\mathrm{s}}{1+\abs{\lambda}^2},
\end{align*}
using the same sign convention as the LHCb experiment~\cite{Aaij:2013oba}. Equation~(\ref{eqnarray:decayrate}) represents the model for \Bs. The model for \Bsbar{} is obtained by changing the sign of the $c_i$ and $d_i$ terms.
The parameters $\abs{A_{\perp}}^2$, $\abs{A_0}^2$, and $\abs{A_{\parallel}}^2$ are the magnitudes squared of the perpendicular, longitudinal, and parallel $P$-wave amplitudes, respectively; $\abs{A_S}^2$ is the magnitude squared of the $S$-wave amplitude representing the fraction of nonresonant decay $\mathrm{B^0_s}\to \JPsi\, \PKp\PKm$; the parameters $\delta_{\perp}$, $\delta_0$, $\delta_{\parallel}$, and $\delta_S$ are their corresponding strong phases.

The complex parameters $\lambda_f$ are defined as $\lambda_f = (q/p) (A_f/\overline{A}_f)$, where the amplitudes $A_f$ ($\overline{A}_f)$ describe the decay of a \Bs{} (\Bsbar{}) meson to a final state $f$, and the parameters $p$ and $q$ relate the mass and flavour eigenstates as $\mathrm{B}_\mathrm{H} = p \mathrm{B^0_s} - q \mathrm{\overline{B}{}^0_s}$ and $\mathrm{B}_\mathrm{L} = p \mathrm{B^0_s} + q \mathrm{\overline{B}{}^0_s}$~\cite{branco}. Assuming polarisation-independent CP-violation effects, $\lambda_f$ can be simplified as $\lambda_f=\eta_f\lambda$, where $\eta_f$ is the CP eigenvalue.
The amount of CP violation in mixing is assumed to be negligible~\cite{Aaij:2013gta}. Thus, no $\abs{q/p}$ terms are used in Eq.~(\ref{eqnarray:decayrate}) when going from the \Bs{} model to the \Bsbar{} model.
Since direct CP violation is expected to be small theoretically~\cite{Charles:2011va} and is measured to be small~\cite{Aaij:2014oba},  $\abs{\lambda}$ is set to 1.0.

\section{The CMS detector}

The central feature of the CMS apparatus is a 13\unit{m} long superconducting solenoid of 6\unit{m} internal diameter, providing a magnetic field of 3.8~T. Within the solenoid volume are a silicon pixel and strip tracker, a lead tungstate crystal electromagnetic calorimeter, and a brass and scintillator hadron calorimeter,
each composed of a barrel and two endcap sections. Muons are measured in gas-ionisation detectors embedded in the steel flux-return yoke outside the solenoid. Extensive forward calorimetry complements the coverage provided by the barrel and endcap detectors.

The main subdetectors used for the present analysis are the silicon tracker and the muon detection system. The silicon tracker measures charged particles within the pseudorapidity range $\abs{\eta}< 2.5$. It consists of 66 million $100{\times}150\mum^2$ silicon pixels and more than 9 million silicon strips. For nonisolated particles of transverse momentum $1 < \pt< 10$\GeV and $\abs{\eta} < 1.4$, the track resolutions are typically 1.5\% in \pt and 25--90 (45--150)\mum in the transverse (longitudinal) impact parameter~\cite{Chatrchyan:2014fea}.

Muons are measured in the pseudorapidity range $\abs{\eta}< 2.4$, with detection planes made using three technologies: drift tubes, cathode strip chambers, and resistive plate chambers.
The relative \pt resolution for low transverse momentum muons with $\pt<10$\GeV is between 0.8\% and 3.0\% depending on $\abs{\eta}$~\cite{Chatrchyan:2012xi}.

The first level (L1) of the CMS trigger system, composed of custom hardware processors, uses information from the calorimeters and muon detectors to select the most interesting events in a fixed time interval of less than 4\mus. The high-level trigger (HLT) processor farm further reduces the event rate from around 100\unit{kHz} to around 400\unit{Hz}, before data storage. At the HLT stage there is full access to all the event information, including tracking, and
therefore selections similar to those applied offline can be used.

A more detailed description of the CMS detector, together with a definition of the coordinate system used and the relevant kinematic variables, can be found in Ref.~\cite{Chatrchyan:2008zzk}.

\section{Event selection and simulated samples}

A trigger optimised for the detection of B hadrons decaying to \JPsi{} is used to collect the data sample.
The L1 trigger  used in this analysis requires two muons, each with \pt greater than 3\GeV and $\abs{\eta} < 2.1$.
The HLT requires a \JPsi{} candidate displaced from the luminous region.
Each muon \pt is required to be at least 4\GeV and the \pt of the reconstructed muon pair must be greater than 6.9\GeV. The \JPsi{} candidates are reconstructed from the muon pairs selected by the trigger in the invariant mass window 2.9--3.3\GeV. The three-dimensional (3D) distance of closest approach of the two muons to each other is required to be smaller than 0.5\unit{cm}. The two muon trajectories are fitted to a common decay vertex. The transverse decay length significance $L_{xy}/\sigma_{L_{xy}}$ is required to be greater than 3, where $L_{xy}$ is the distance between the centre of the luminous region and the secondary vertex in the transverse plane, and $\sigma_{L_{xy}}$ is the $L_{xy}$ uncertainty. The secondary-vertex fit probability, calculated using the $\chi^2$ and the number of degrees of freedom of the vertex fit, must be greater than 10\%.  The angle $\rho$ between the dimuon transverse momentum and the $L_{xy}$ direction is required to satisfy $\cos\rho >0.9$.

Offline selection criteria are applied to the sample.
The individual muon candidates are required to lie within a kinematic acceptance region of $\pt> 4$\GeV and
$\abs{\eta}<2.1$.
Two oppositely charged muon candidates are paired and required to originate from a common vertex.
Dimuon candidates with invariant mass within 150 \MeV{} of the world-average \JPsi{} mass~\cite{Agashe:2014kda} are selected.
Candidate $\phi(1020)$ mesons are reconstructed from pairs of oppositely charged tracks with $p_{T} > 0.7$\GeV, after removing the muon candidate tracks forming the \JPsi. Each selected track is assumed to be a kaon, and the invariant mass of a track pair is required to be within \mbox{10~\MeV}~of the world-average $\phi(1020)$ mass~\cite{Agashe:2014kda}.

The \Bs{} candidates are formed by combining \JPsi{} and $\phi$(1020) candidates.
A kinematic fit of the two muon and two kaon candidates is performed, with a common vertex, and the dimuon invariant mass is constrained to the nominal \JPsi{} mass~\cite{Agashe:2014kda}.
A \Bs{} candidate is retained if the $\JPsi{}\,\phi(1020)$ pair has an invariant mass between 5.20 and 5.65\GeV and the $\chi^2$ vertex fit probability is greater than 2\%.

Multiple pp collisions can occur in the same beam crossing (pileup).
The average number of primary vertices in an event is approximately 16, and each selected event is required to have at least one reconstructed primary vertex. If there are multiple vertices, the one that minimises the angle between the flight direction and the momentum of the \Bs{} is selected. The selected primary vertex is used to measure $ct$. The quantity $ct$ is calculated  from the transverse decay length vector of the \Bs{}, $\vec{\textit{L}}_{xy}^{\mathrm{B^0_s}}$, as $ct = m_\mathrm{PDG}^{\mathrm{B^0_s}} \vec{\textit{L}}_{xy}^{\mathrm{B^0_s}} \cdot \vec{\textit{p}}_{\mathrm{T}}/ \pt^2$, where $m_\mathrm{PDG}^{\mathrm{B^0_s}}$ is the world-average \Bs{} mass~\cite{Agashe:2014kda} and $\vec{\textit{p}}_{\mathrm{T}}$ is the \Bs{} transverse momentum vector. The decay length is calculated in the transverse plane to minimise effects due to pileup.

Simulated events are produced using the \PYTHIA v6.424 Monte Carlo event generator~\cite{Sjostrand:2006za}. The B hadron decays are modelled with the \EVTGEN simulation package~\cite{Lange2001152}. For the \Bs{} signal generation, the \textsc{evtpvvcplh} module is used, which simulates the double vector decay taking into account neutral meson mixing and  CP-violating time-dependent asymmetries. Final-state radiation is included in \EVTGEN through the \PHOTOS package~\cite{Barberio1991115, Barberio1994291}. The events are then passed through a detailed
\GEANTfour-based simulation~\cite{Agostinelli2003250} of the CMS detector.
The predicted distributions from simulation of many kinematic and geometric variables are compared to those from data and found to be in agreement.
The simulated samples are used to determine the signal reconstruction efficiencies, and to study the background components in the \Bs{} signal mass window.

\begin{figure}[h!t]
  \centering
    \includegraphics[width=\cmsFigWidth]{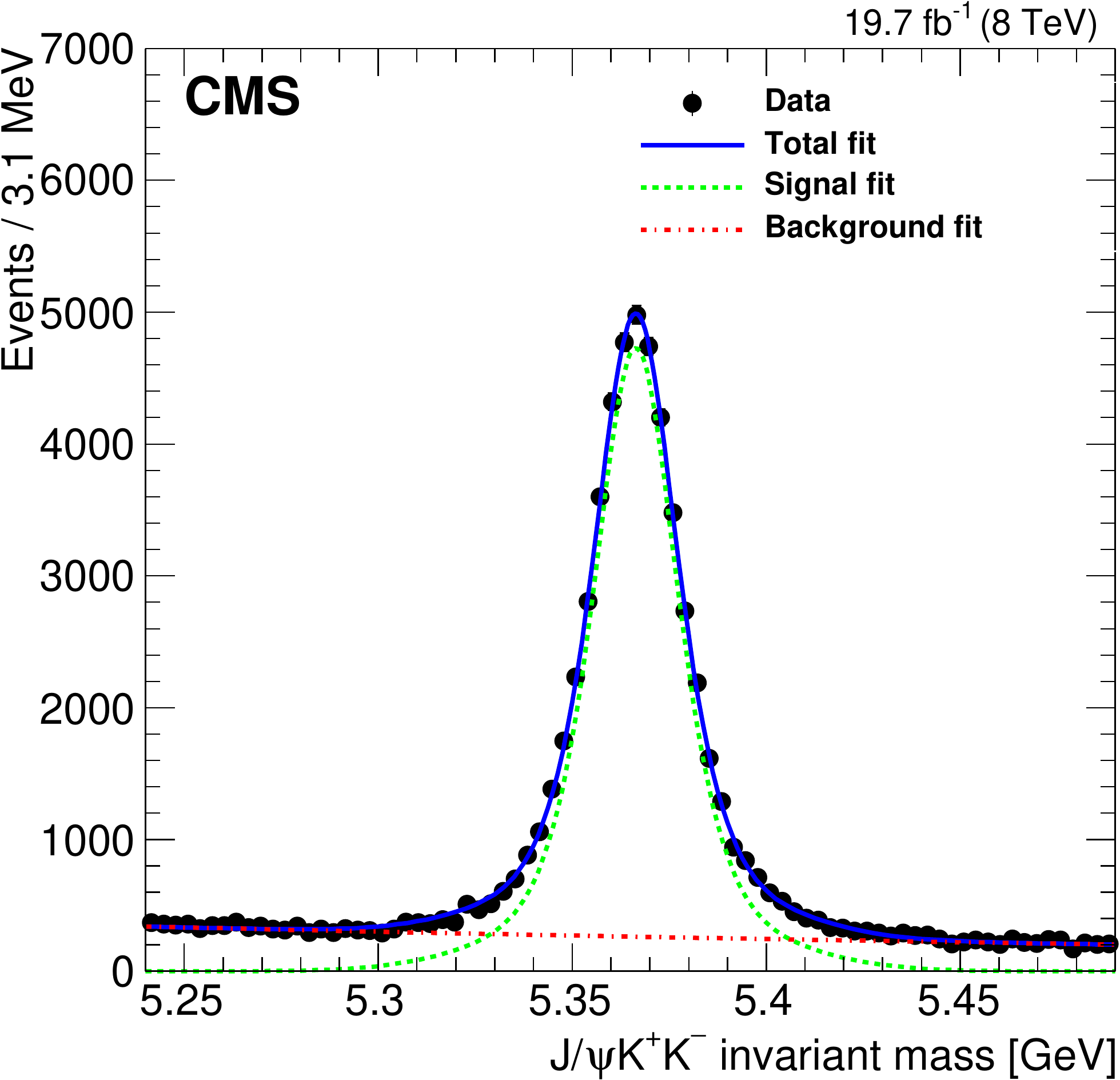}
    \caption{The $\JPsi\PKp\PKm$ invariant mass distribution of the \Bs{} candidates. The solid line is a fit to the data (solid markers), the dashed line is the signal component and the dot-dashed line is the background component.}
    \label{fig:mass}
\end{figure}

\begin{figure}[h!]
\centering
    \includegraphics[width=0.48\textwidth]{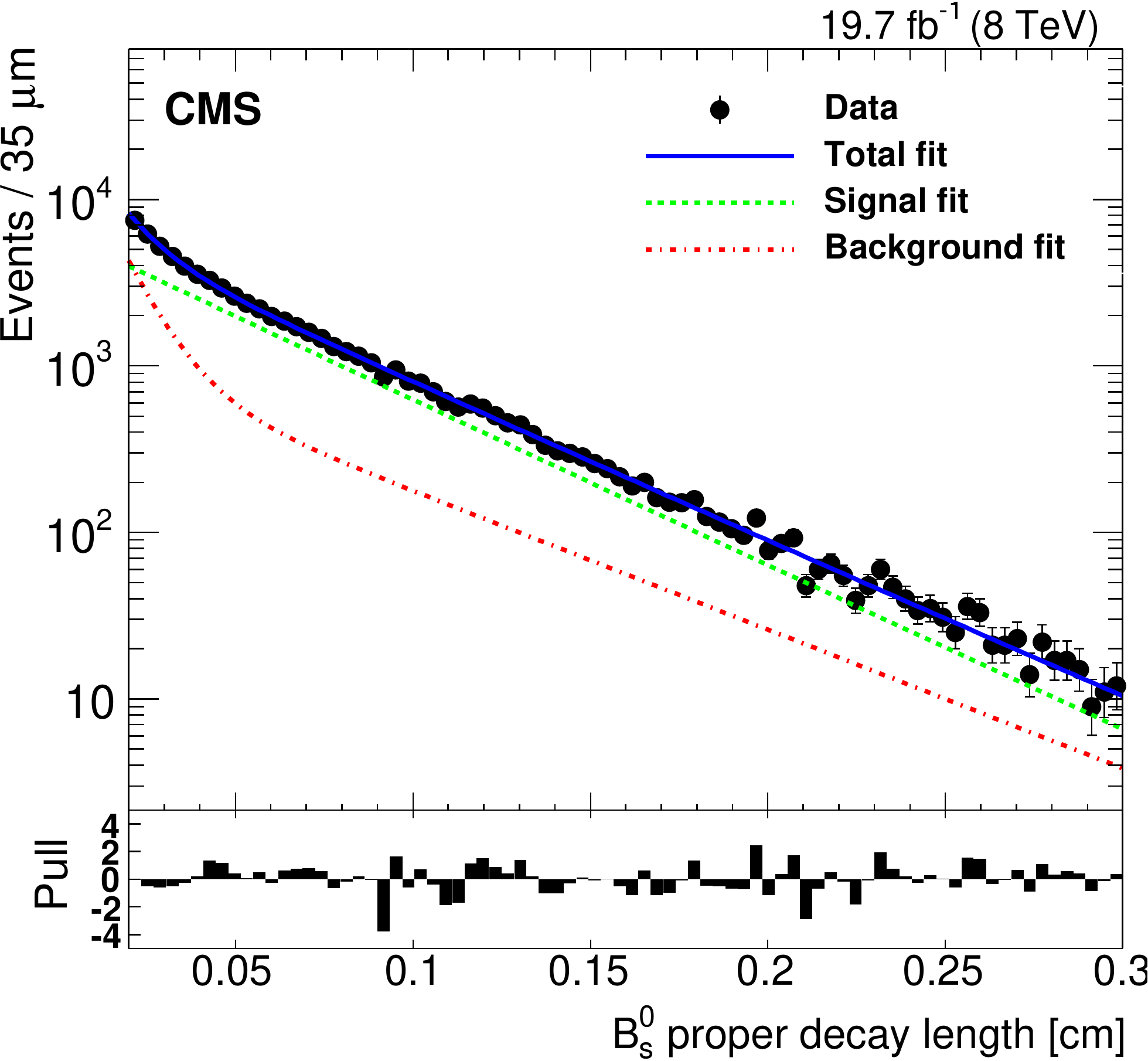}
    \includegraphics[width=0.48\textwidth]{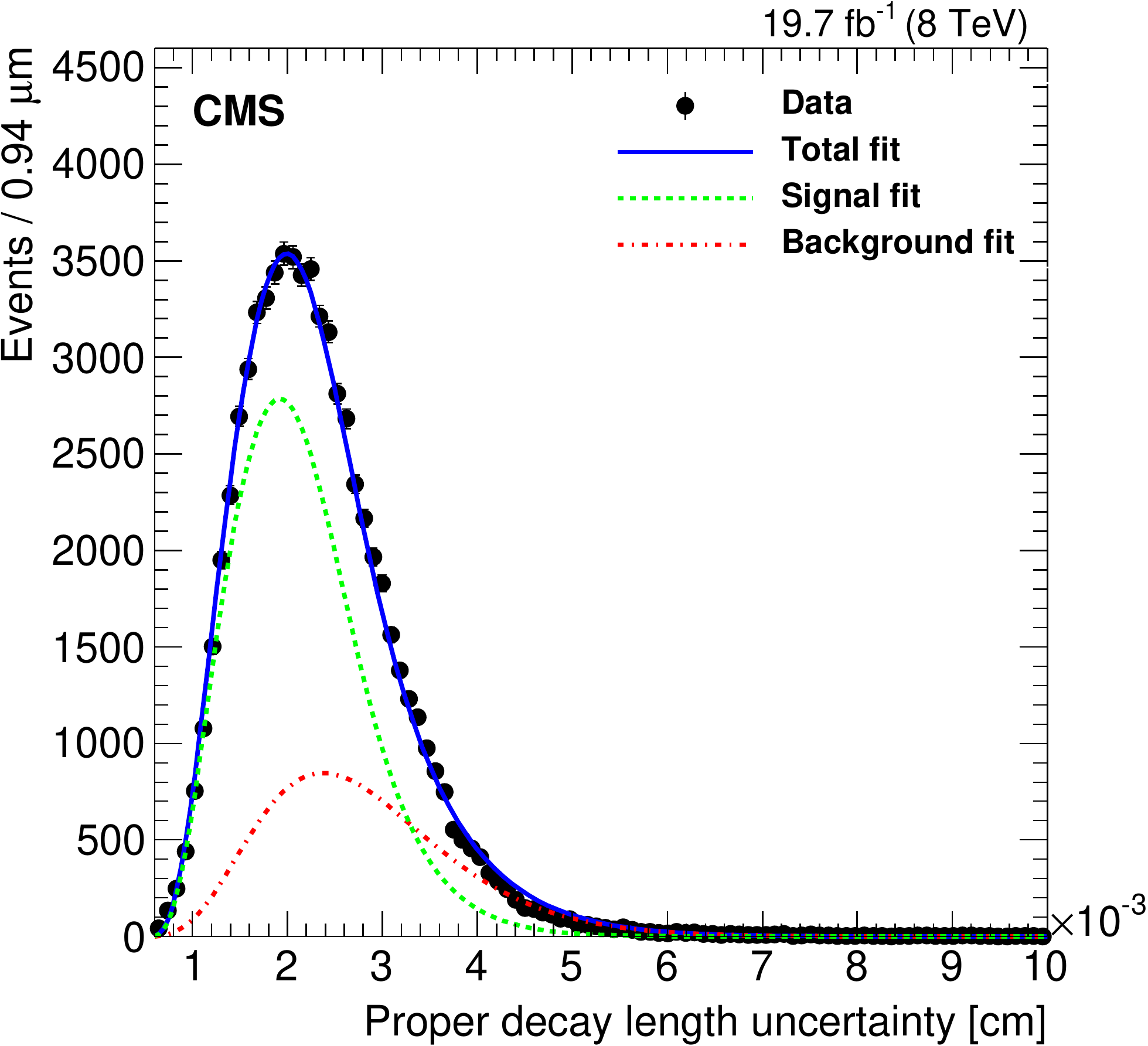}
    \caption{The $ct$ distribution (\cmsLeft) and its uncertainty $\sigma_{ct}$ (\cmsRight) of the \Bs{} candidates. The solid line is a fit to the data (solid markers), the dashed line is the signal component and the dot-dashed line is the background component. For the $ct$ distribution the pull, defined as the difference between the observed events and the fit function applied to the sum of the signal and background, divided by the statistical uncertainty in the observed events, is displayed in the histogram in the lower panel.}
    \label{fig:lifetime}
\end{figure}

The main background for the \BsJpsiPhi{} decays originates from nonprompt \JPsi{} mesons from the decay of B hadrons, such as \Bd, \Bu, \Lb, and \Bc.
Since the \Bc{} cross section is small~\cite{Agashe:2014kda} compared to that of the \Bs{}~\cite{Agashe:2014kda}, the \Bc{} decays can be neglected.
The contribution of the $\Lambda_\mathrm{b} \to \JPsi \mathrm{X}$ channels to the selected events is also found to be small, and its mass distribution in the selected mass range is observed to be flat.
The effect of background with a similar signal signature on the physics observables is studied using simulated events, and found to be negligible. The mass distribution in the signal region is shown in Fig.~\ref{fig:mass}, and the distribution of $ct$  and its uncertainty $\sigma_{ct}$ in Fig.~\ref{fig:lifetime}.

Efficiency corrections owing to the detector acceptance, trigger selection, and selection criteria applied in the data analysis are taken into account in the modelling of the angular observables. The angular efficiency $\epsilon(\Theta)$ is calculated using a fully simulated sample of $\BsJpsiPhi{}\to\mu^+\mu^-\PKp\PKm$ decays. In this sample, the $\Delta\Gamma_{\mathrm{s}}$ parameter is set to zero to avoid correlations between the decay time and the angular variables. The $\epsilon(\Theta)$ is fitted to a 3D function of $\Theta$ to properly account for the correlation among the angular observables.

The trigger includes a decay length significance requirement for the \JPsi{} candidates. Accordingly, the value of $ct$ is required to be greater than 200~$\mu$m in order to avoid a lifetime bias coming from the turn-on curve of the trigger efficiency.
The efficiency histogram of $ct$ is then fitted with a straight line plus a sigmoid function.

\section{Flavour tagging}

The flavour of each \Bs{} candidate at production time is determined with an opposite-side (OS) flavour tagging algorithm.
Since b quarks are produced as ${\mathrm{b\bar{b}}}$ pairs, the flavour of the signal \Bs{} meson at production time can be inferred from the flavour of the other B hadron in the event.
The tagging algorithm used in this analysis requires an additional muon or electron in the events containing a reconstructed ${\mathrm{B^0_s}\to \JPsi\,\phi(1020)}$ decay. The additional lepton
is assumed to originate from a semileptonic decay of the OS B hadron, $\mathrm{b}\to\ell\nu\mathrm{X}$ decay, with $\ell = \Pe,\mu$.
For all the events in which an OS tag lepton is found the algorithm provides a tag decision $\xi$ based on the charge of the lepton: $\xi=+1$ for signal \Bs, and $\xi=-1$ for signal \Bsbar.

The tag decision is affected by processes that reverse the charge-flavour correlation, such as cascade decays $\PQb\to \PQc\to \ell$, or semileptonic decays of neutral OS $\mathrm{B}$ mesons that have oscillated to their antiparticles before decaying. Leptons produced from flavour-uncorrelated sources, such as semileptonic decays of promptly produced charmed hadrons, pion and kaon decays, $\JPsi\,$ decays, and Dalitz decays of neutral pions further contribute to diluting the tag information. The probability of assigning a wrong flavour to the signal \Bs{} is described by the mistag probability $\omega$, defined as the ratio of the number of wrongly tagged events divided by the total number of tagged events, which is directly related to the dilution factor $D = \left(1-2\omega\right)$. The value of $\omega$ is estimated from data on a per-event basis, as described below.

The tagging algorithm is optimised by maximising the tagging power $\mathcal{P}_\text{tag} = \varepsilon_\text{tag}(1-2\omega)^2$, which represents the equivalent efficiency of a sample with perfect tagging ($\omega=0$).
The term $\varepsilon_\text{tag}$ is the tagging efficiency, defined as the fraction of events to which a tag decision is found by the tagging algorithm.

Opposite-side muons and electrons are reconstructed with the particle-flow algorithm~\cite{CMS:2009nxa, CMS:2010byl}.
In each event, the muons (electrons) that are not part of the reconstructed \BsJpsiPhi{} decay are required to be identified with loose identification criteria.
If there are multiple muons (electrons) in the event, the one with the highest \pt is chosen at this stage.
The tag lepton selections are then optimised separately for muons and electrons using simulated signal samples of ${\mathrm{B^0_s}\to \JPsi\,\phi(1020)}$ decays.
A cut-based opposite-side lepton selection is applied to reduce the number of leptons not originating from B-hadron decays. To optimise the selection, several variables are studied, and a set of five discriminating variables ($p_\mathrm{T}$, $\eta$, $d_{xyz}$, $\Delta\mathrm{R}$, Isolation) is identified. A total number of more than four million alternative cut configurations have been tested to determine the configuration that maximises the tagging power, independently for muons and electrons.
The tag muon is thus required to have $\pt>2.2\GeV$, the 3D impact parameter $d_{xyz}$ with respect to the primary vertex associated with the signal \Bs{} is required to be smaller than 0.1\unit{cm}, and the angular separation, $\Delta R = \sqrt{\smash[b]{(\Delta \phi)^2 + (\Delta \eta)^2}}$, between the muon and the signal \Bs{} is required to be greater than 0.3, where $\Delta\phi$ and $\Delta\eta$ are the azimuthal angle and pseudorapidity differences between the directions of the tag muon and the \Bs{} candidate.
Electrons are required to have $\pt> 2.0\GeV$, $d_{xyz} <0.1\unit{cm}$, and $\Delta R  >0.2$.
In addition, a multivariate discriminator ($\mathrm{MVA}_{e-\pi}$) tuned to separate genuine electrons reconstructed by the particle-flow algorithm from pions and photons is applied to tag electrons by requiring that the discriminator is greater than 0.2~\cite{CMS:2010byl}.

A multilayer perceptron neural network (MLP-NN) of the TMVA toolkit~\cite{Hocker:2007ht} is used to further separate the right- and wrong-tag leptons.
Training and testing is performed using approximately 24\,000 and 20\,400 simulated ${\mathrm{B^0_s}\to \JPsi\,\phi(1020)}$ events for the muon and electron MLP-NNs, respectively, and two independently optimised sets of variables. Half of each sample is used for training and the other half for testing. The input variables common to both MLP-NNs are \pt,  $\eta$, and $d_{xyz}$ of the tag lepton, and two variables related to activity in a cone around the lepton direction: a particle-flow relative isolation variable~\cite{CMS:2010byl} and a \pt-weighted average of the charges of the particles in the cone. Specific variables are further introduced in the MLP-NNs separately for muons and electrons. For muons, the \pt relative to the axis of the jet containing the muon is used, while for electrons the $\mathrm{MVA}_{e-\pi}$ is exploited.

The mistag probabilities are obtained from data using the self-tagging channel $\mathrm{B^\pm}\to \JPsi\,\mathrm{K^\pm}$, where the charge of the reconstructed kaon determines the flavour of the $\mathrm{B^\pm}$ and, in the absence of mixing, of the opposite-side B hadron as well.
The mistag probabilities are parametrised separately for muons and electrons with analytic functions of the MLP-NN discriminators in order to provide a per-event value of the predicted mistag probability $\omega$.
The functional forms of the parametrisations are obtained from the simulated \Bs{} sample.
The candidate \Bu{} mesons are required to pass a selection as similar as possible to that applied for the reconstruction of the signal \Bs{} candidates.
The same trigger and $\JPsi$ reconstruction requirements as for the \Bs{} signal sample are applied. A charged particle with $\pt> 2\GeV$, assumed to be a kaon, is combined with the dimuon pair in a kinematic fit.
An unbinned extended maximum-likelihood fit to the invariant $\JPsi\,\mathrm{K}^\pm$ mass is performed, yielding a total of $(707\pm2)\times10^3$ reconstructed $\mathrm{B^\pm}\to \JPsi\,\mathrm{K^\pm}$ events.
The tagging efficiency evaluated with the $\mathrm{B^\pm}\to \JPsi\,\mathrm{K^\pm}$ data sample is $(4.56\pm0.02)\%$ and $(3.92\pm0.02)\%$ for muons and electrons, respectively, where the uncertainties are statistical.

\begin{figure}[!htbp]
 \centering
  \includegraphics[width=0.47\textwidth]{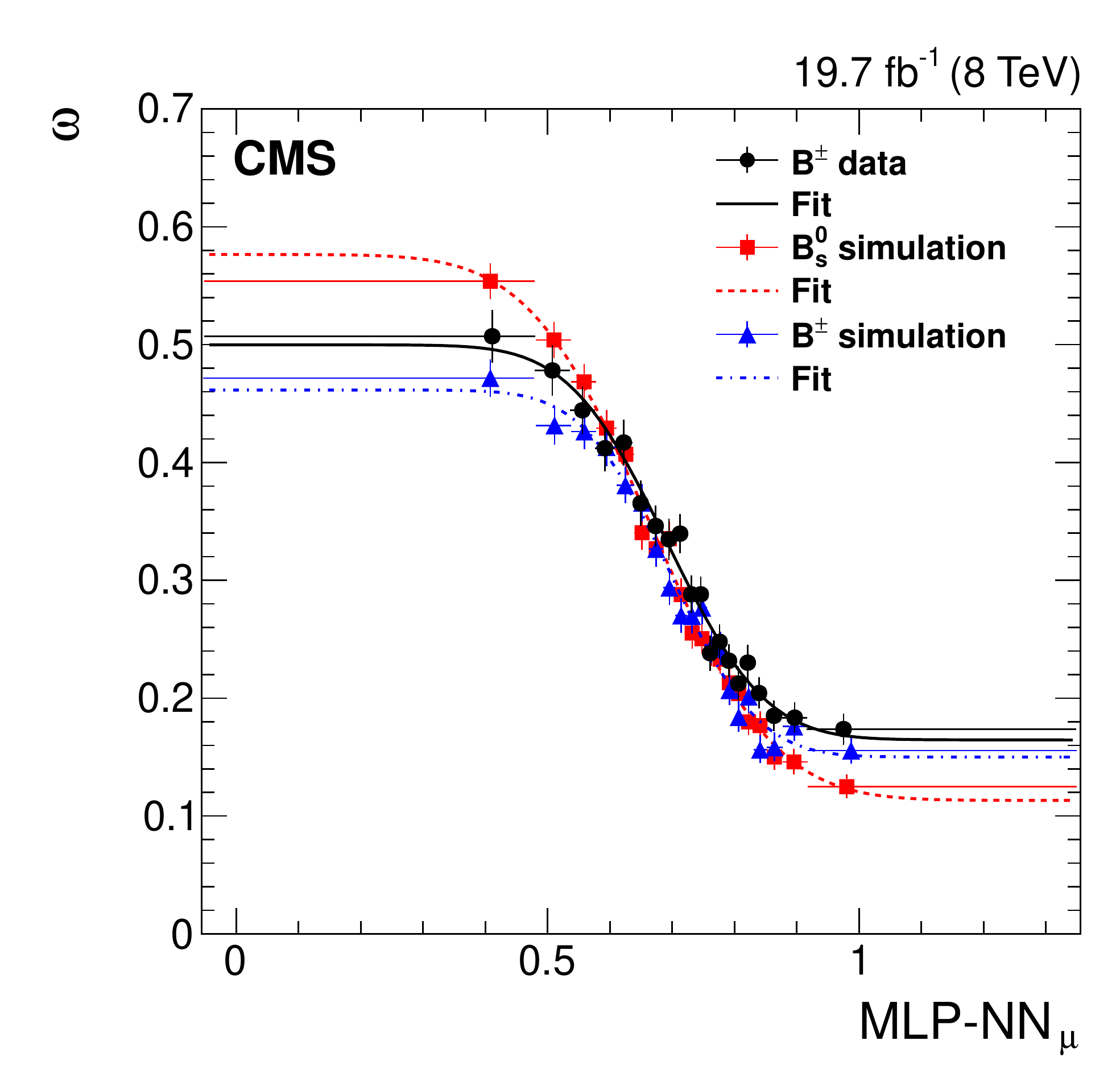}
  \includegraphics[width=0.47\textwidth]{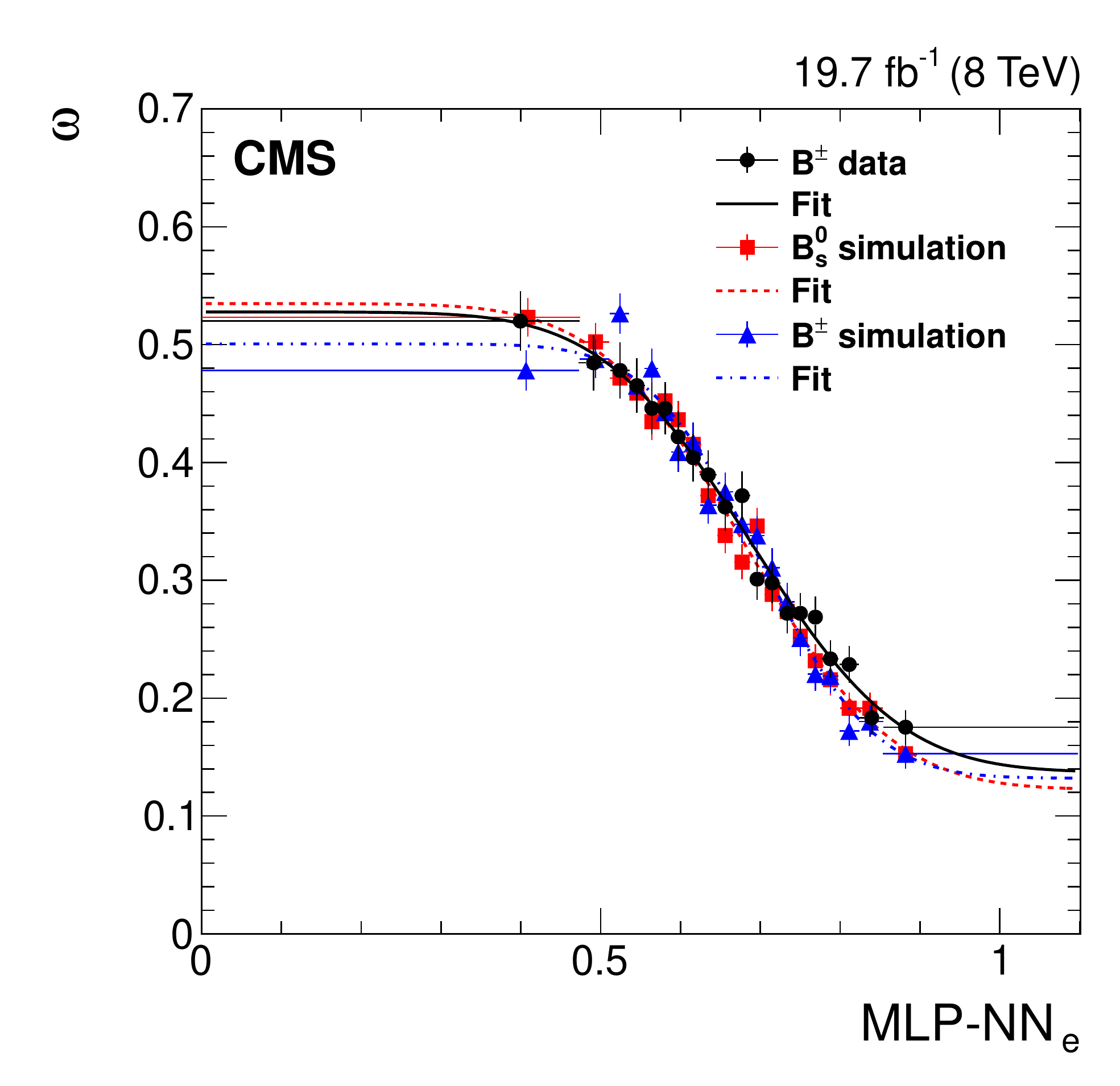}
 \caption{
The mistag probabilities $\omega$, defined as the ratio of the number of wrongly tagged events divided by the total number of tagged events, as a function of the MLP-NN discriminators for muons (\cmsLeft) and electrons (\cmsRight). The data points (solid markers) are placed at the average weighted value of the events in each bin. The vertical bars show the statistical uncertainties and the horizontal bars the bin width. The solid line represents the parametrisation curve extracted from the background-subtracted \Bu{} data; the dashed and dot-dashed lines refer to the parametrisations extracted from the simulated \Bs{} and \Bu{} samples, respectively.
 \label{fig:parametrisation}}
\end{figure}

The mistag parametrisation curves evaluated with the \Bu{} control channel for muons and electrons are shown in Fig.~\ref{fig:parametrisation}, where the parametrisations for the \Bu{} and \Bs{} simulated samples are shown for comparison.

Most tagged events have only a single electron or muon tag.
If both tags are available for a specific event (about 3.5\% of the cases), the tag lepton with the greatest value of the dilution factor is retained, and the tag decision and the estimated mistag are taken from this tag lepton.
The overall lepton tagging efficiency is $(8.31\pm0.03)\%$, as measured in data with the $\mathrm{B^\pm}\to \JPsi\,\mathrm{K^\pm}$ data sample.

To correct for potential effects induced by the dependence of the tagging algorithm on the ${\mathrm{B^0_s}\to \JPsi\,\phi(1020)}$ simulation, the mistag probability is calibrated by comparing the per-event predicted $\omega$ to the measured $\omega^\text{meas}$ obtained from the $\mathrm{B^\pm}\to \JPsi\,\mathrm{K^\pm}$ data control channel.
This is then fit to the function $\omega^\text{meas} = p_0 + p_1(\omega-\omega')$, chosen to limit the correlation between the function parameters $p_0$ and $p_1$. The parameter $\omega'$ is fixed to a value roughly corresponding to the mean of the calculated mistag probability, $\omega'=0.35$.
The resulting calibration parameters are $p_0=0.348\pm0.003$ and $p_1=1.01\pm0.03$, and their uncertainties are propagated as a statistical uncertainty in the OS tagger.

The systematic uncertainties related to the calibration parameters $p_0$ and $p_1$ are dominated by the dependence of these parameters on the flavour of the signal-side B hadron. The uncertainties are estimated from \Bu{} data and simulated samples of \Bs{} and \Bu{} events. Systematic uncertainties originating from possible variations in the CMS data-taking conditions, the signal B hadron kinematics, the analytic form of the mistag parametrisation functions, and the model used to fit the \Bu{} invariant mass distribution are tested and found to be negligible.

The overall tagging power of the OS lepton tagger, measured with a sample of $\mathrm{B^\pm}\to \JPsi\,\mathrm{K^\pm}$ events, is $\mathcal{P}_\text{tag}=\left(1.307\pm0.031\stat\pm0.007\syst\right)\%$, corresponding to the combined mistag probability $\omega=\left(30.17\pm0.24\stat\pm0.05\syst\right)\%$.

\section{Maximum-likelihood fit}

An unbinned maximum-likelihood fit to the data is performed by including the information on the \Bs{} invariant mass ($m_{\mathrm{B_s^0}}$), the three decay angles  ($\Theta$) of the reconstructed \Bs{} candidates, the flavour tag decision ($\xi$), $ct$, and $\sigma_{ct}$, obtained by summing in quadrature the decay length uncertainty and the uncertainty in the transverse momentum. The fit is applied to the sample of $70\,500$ events, out of which $5\,650$ are tagged events, selected in the mass range 5.24--5.49\GeV and  $ct = 200$--3\,000\mum. From this multidimensional fit, the physics parameters of interest \deltagammas{}, $\phi_\mathrm{s}$, the \Bs{} mean lifetime $c\tau$, $\abs{A_{\perp}}^2$, $\abs{A_{0}}^2$, $\abs{A_S}^2$, and the strong phases $\delta_{\parallel}$, $\delta_{\perp}$, and $\delta_{S\perp}$ are determined, where $\delta_{S\perp}$ is defined as the difference $\delta_{S}-\delta_{\perp}$. The P-wave amplitudes are normalised to unity by constraining $\abs{A_{\parallel}}^2$ to $1-\abs{A_{\perp}}^2-\abs{A_{0}}^2$. The fit model is validated with simulated pseudo-experiments and with simulated samples with different parameter sets.

The likelihood function is composed of probability density functions (pdf) describing the signal and background components.
The likelihood fit algorithm is implemented using the RooFit package from the ROOT framework~\cite{Antcheva20092499}. The signal and background pdfs are formed as the product of pdfs that model the invariant mass distribution and the time-dependent decay rates of the reconstructed candidates. In addition, the signal pdf also includes the efficiency function. The event likelihood function ${\cal{L}}$ is represented as:
\ifthenelse{\boolean{cms@external}}{
\begin{equation*}
\begin{split}
{\mathcal{L}} =& L_\mathrm{s} +  L_\text{bkg}, \\
L_{\text{s}} =&  N_\mathrm{s}\, \left[\tilde{f}\left(\Theta,ct,\alpha\right) \otimes G\left(ct,\sigma_{ct}\right)\, \epsilon\left(\Theta\right) \right] \\ &\times P_\mathrm{s}(m_{\mathrm{B_s^0}})\, P_\mathrm{s}(\sigma_{ct})\, P_\mathrm{s}(\xi), \\
L_{\text{bkg}}  =&  N_\text{bkg}\, P_\text{bkg}(\cos\theta_\mathrm{T},\varphi_\mathrm{T})\, P_\text{bkg}(\cos\psi_\mathrm{T})\, P_\text{bkg}(ct) \\ &\times P_\text{bkg}(m_{\mathrm{B_s^0}}) \, P_\text{bkg}(\sigma_{ct})\, P_\text{bkg}(\xi),
\label{eqt:Fullpdf}
\end{split}
\end{equation*}
}{
\begin{equation*}
\begin{split}
{\mathcal{L}} &= L_\mathrm{s} +  L_\text{bkg}, \\
L_{\text{s}} &=  N_\mathrm{s}\, \left[\tilde{f}\left(\Theta,ct,\alpha\right) \otimes G\left(ct,\sigma_{ct}\right)\, \epsilon\left(\Theta\right) \right] \, P_\mathrm{s}(m_{\mathrm{B_s^0}})\, P_\mathrm{s}(\sigma_{ct})\, P_\mathrm{s}(\xi), \\
L_{\text{bkg}}  &=  N_\text{bkg}\, P_\text{bkg}(\cos\theta_\mathrm{T},\varphi_\mathrm{T})\, P_\text{bkg}(\cos\psi_\mathrm{T})\, P_\text{bkg}(ct) \, P_\text{bkg}(m_{\mathrm{B_s^0}}) \, P_\text{bkg}(\sigma_{ct})\, P_\text{bkg}(\xi),
\label{eqt:Fullpdf}
\end{split}
\end{equation*}
}
where $L_\mathrm{s}$ and $L_\text{bkg}$ are the pdfs that describe the \BsJpsiPhi{} signal and background contributions, respectively. The number of signal (background) events is $N_\mathrm{s}$ ($N_\text{bkg}$). The pdf  $\tilde{f}(\Theta,ct,\alpha)$ is the differential decay rate function $f(\Theta,ct,\alpha)$ defined in Eq.~(\ref{eqnarray:decayrate}), modified to include the flavour tagging information and the dilution term $(1-2\omega)$, which are applied to each of the $c_i$ and $d_i$ terms of the equation. In the $\tilde{f}$ expression, the value of $\delta_0$ is set to zero, following a general convention.
The function $\epsilon(\Theta)$ is the angular efficiency and $G(ct,\sigma_{ct})$ is a Gaussian resolution function, which makes use of the event-by-event decay time uncertainty $\sigma_{ct}$, scaled by a factor $\kappa$. The $\kappa$ factor is a function of $ct$ and is introduced as a correction to take care of residual effects when the decay time uncertainty is used to model the $ct$ resolution. The function $\kappa(ct)$ is measured using simulated samples and, on average, its value equals 1.0 to within a few percent. The average decay time uncertainty including the $\kappa(ct)$ factor equals 23.4 $\mu$m. All the parameters of the pdfs are left free to float in the final fit, unless explicitly stated otherwise. The value of $\Delta\Gamma_\mathrm{s}$ is constrained to be positive, based on recent measurements~\cite{PhysRevLett.108.241801}.

The signal mass pdf $P_\mathrm{s}(m_{\mathrm{B_s^0}})$ is the sum of three Gaussian functions with a common mean; the two smaller widths, the mean, and the fraction of each Gaussian function are fixed to the values obtained in a one-dimensional mass fit. The background mass distribution $P_\text{bkg}(m_{\mathrm{B_s^0}})$ is described by an exponential function. The background  decay time component $P_\text{bkg}(ct)$ is described by the sum of two exponential functions. The angular parts of the backgrounds pdfs $P_\text{bkg}(\cos\theta_\mathrm{T},\varphi_\mathrm{T})$ and  $P_\text{bkg}(\cos\psi_\mathrm{T})$ are described analytically by a series of Legendre polynomials for $\cos\theta_\mathrm{T}$ and $\cos\psi_\mathrm{T}$ and sinusoidal functions for $\varphi_\mathrm{T}$.  For the $\cos\theta_\mathrm{T}$ and $\varphi_\mathrm{T}$ variables a two-dimensional pdf is used to take into account the correlation among the variables.

The signal decay time uncertainty pdf $P_\mathrm{s}(\sigma_{ct})$ is a sum of two Gamma functions, with all the parameters  fixed to the values obtained by fitting a sample of background-subtracted events. The background decay time uncertainty pdf $P_\text{bkg}(\sigma_{ct})$ is represented by a Gamma function. All the parameters are fixed to the values obtained by fitting the \Bs{} invariant mass sideband regions, defined by the mass ranges  $m_{\mathrm{B_s^0}}$ = 5.24--5.28\GeV and 5.45--5.49 \GeV.
The functions $P_\mathrm{s}(\xi)$ and $P_\text{bkg}(\xi)$ are the tag decision $\xi$ pdfs, which have been obtained from the data.

\section{Results and systematic uncertainties}

\begin{figure*}[h!t]
\includegraphics[width=0.31\textwidth]{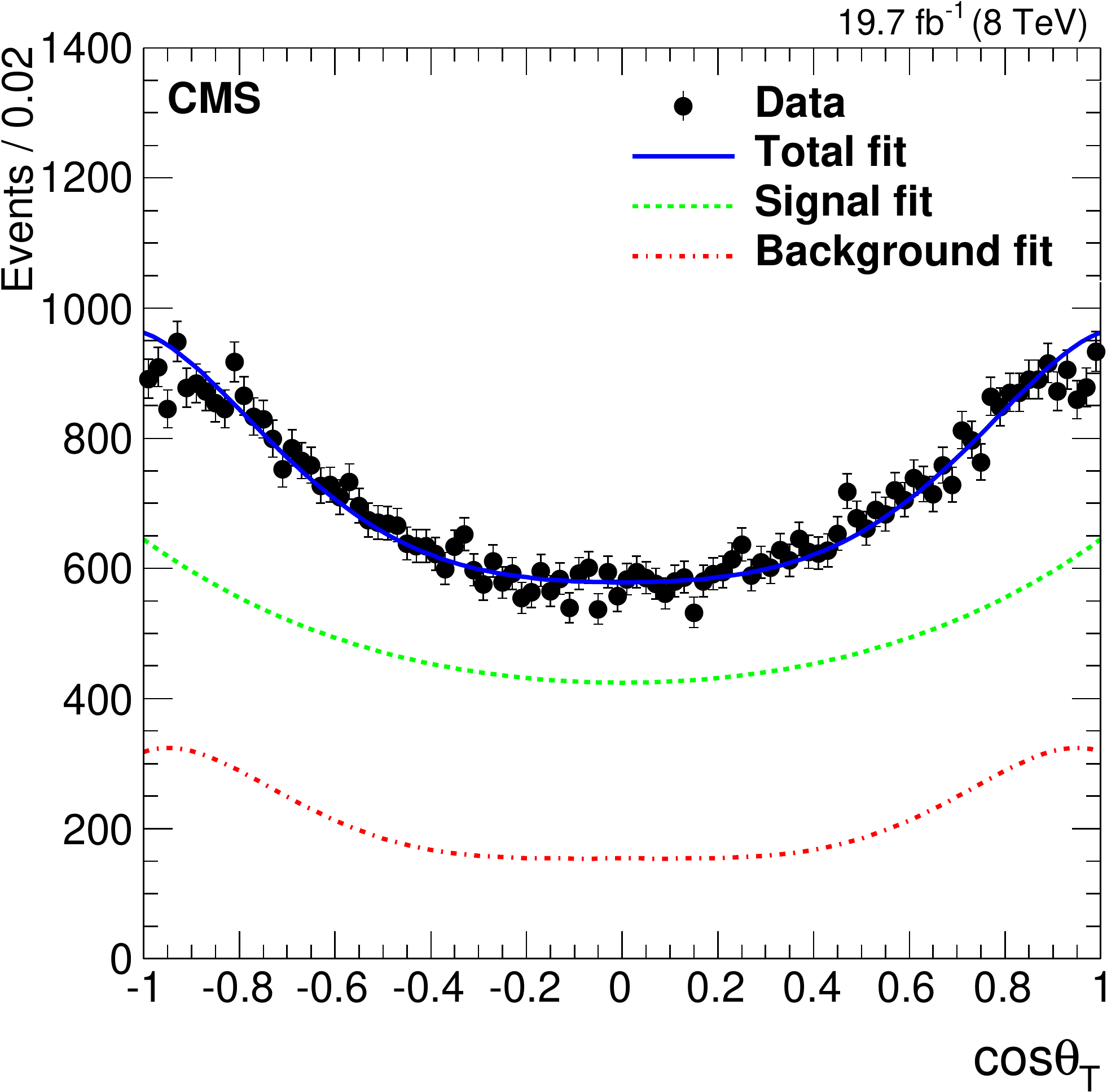} \hfil
\includegraphics[width=0.31\textwidth]{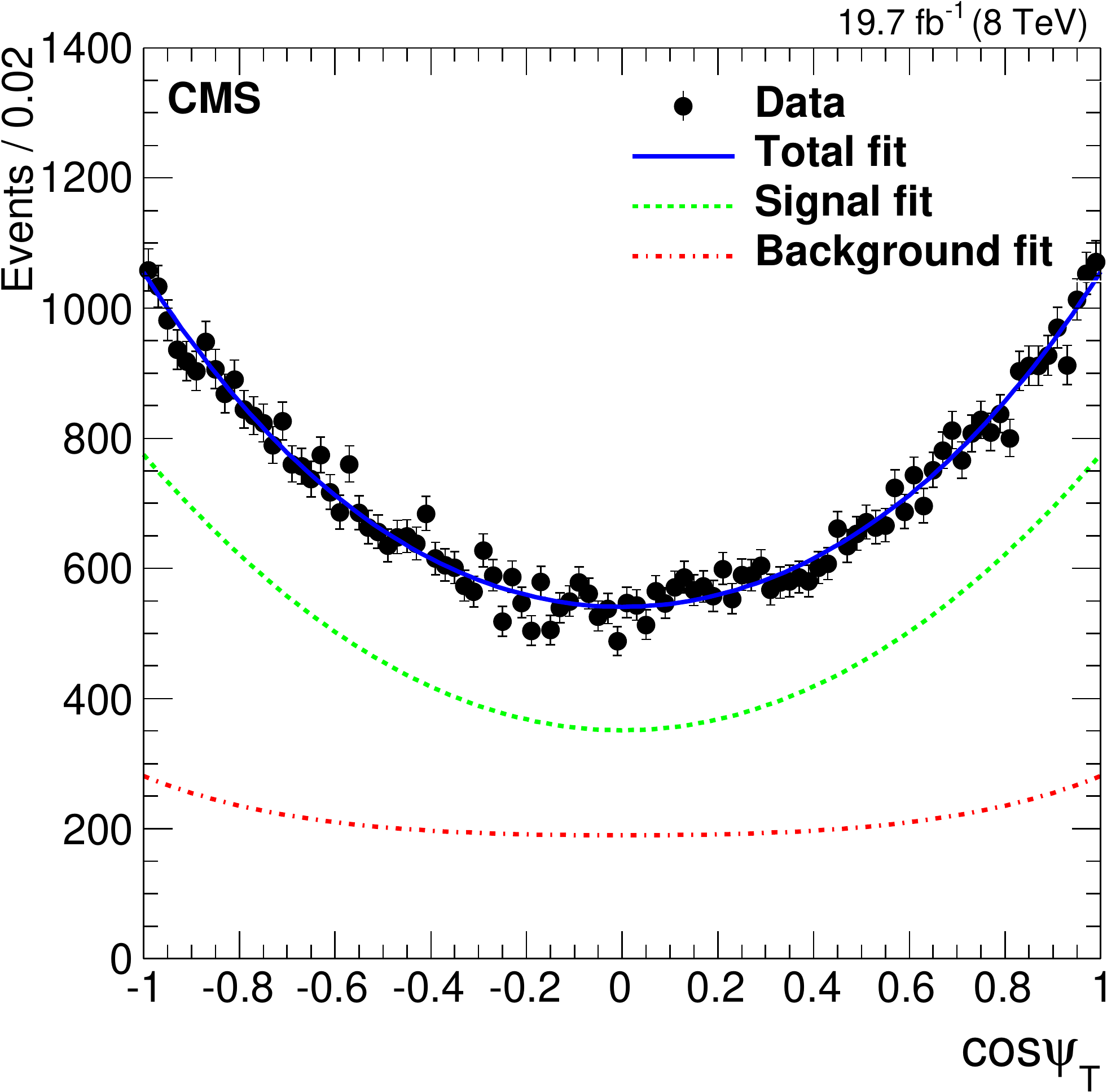} \hfil
\includegraphics[width=0.31\textwidth]{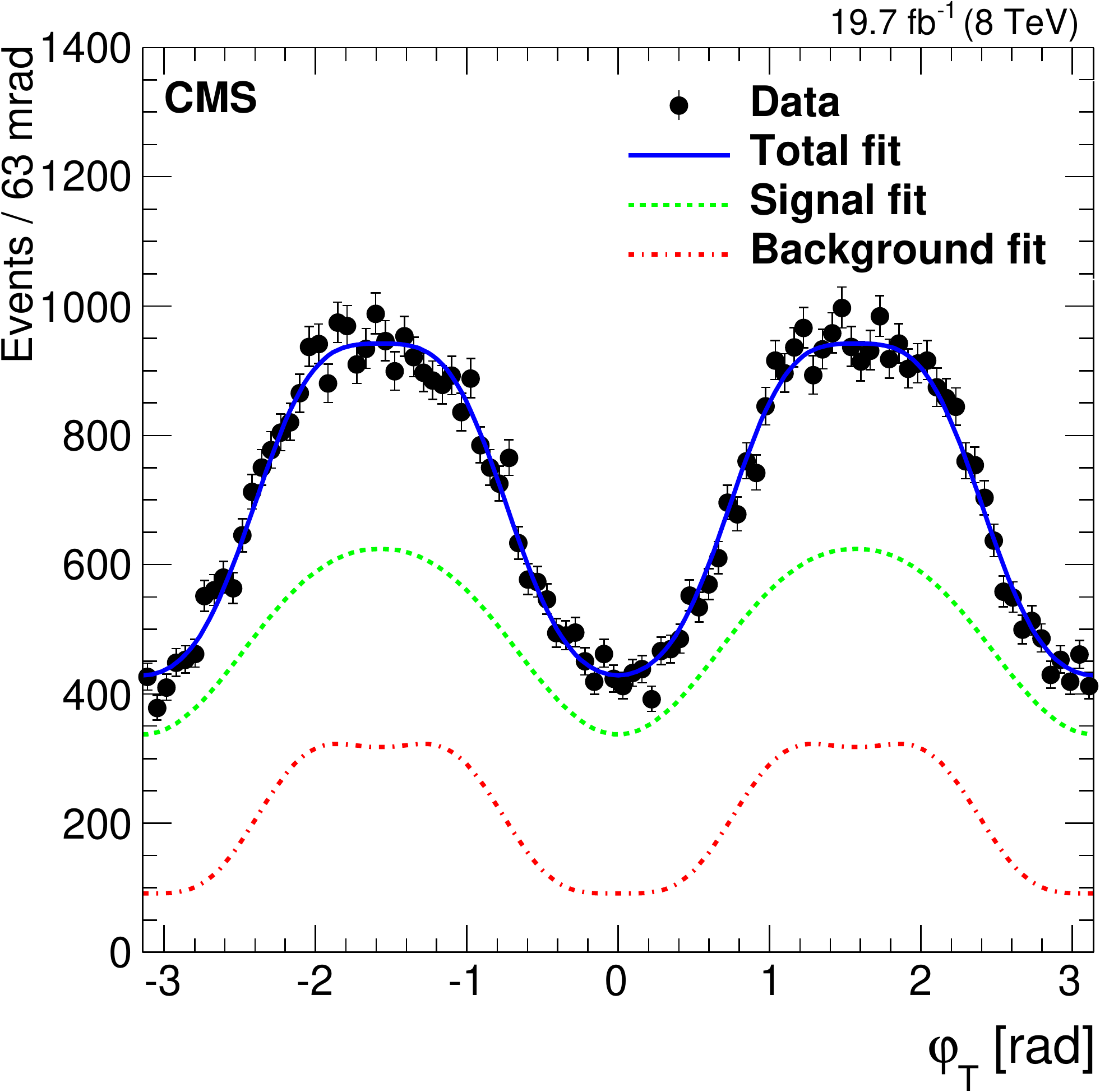}
\caption{The angular distributions ($\cos\theta_\mathrm{T}$, $\cos\psi_\mathrm{T}$, $\varphi_\mathrm{T}$) of the \Bs{} candidates from data (solid markers). The solid line is the result of the fit, the dashed line is the signal component, and the dot-dashed line is the background component.}
\label{fig:datafittagged}
\end{figure*}

\begin{figure}[h!t]
\centering
\includegraphics[width=\cmsFigWidth]{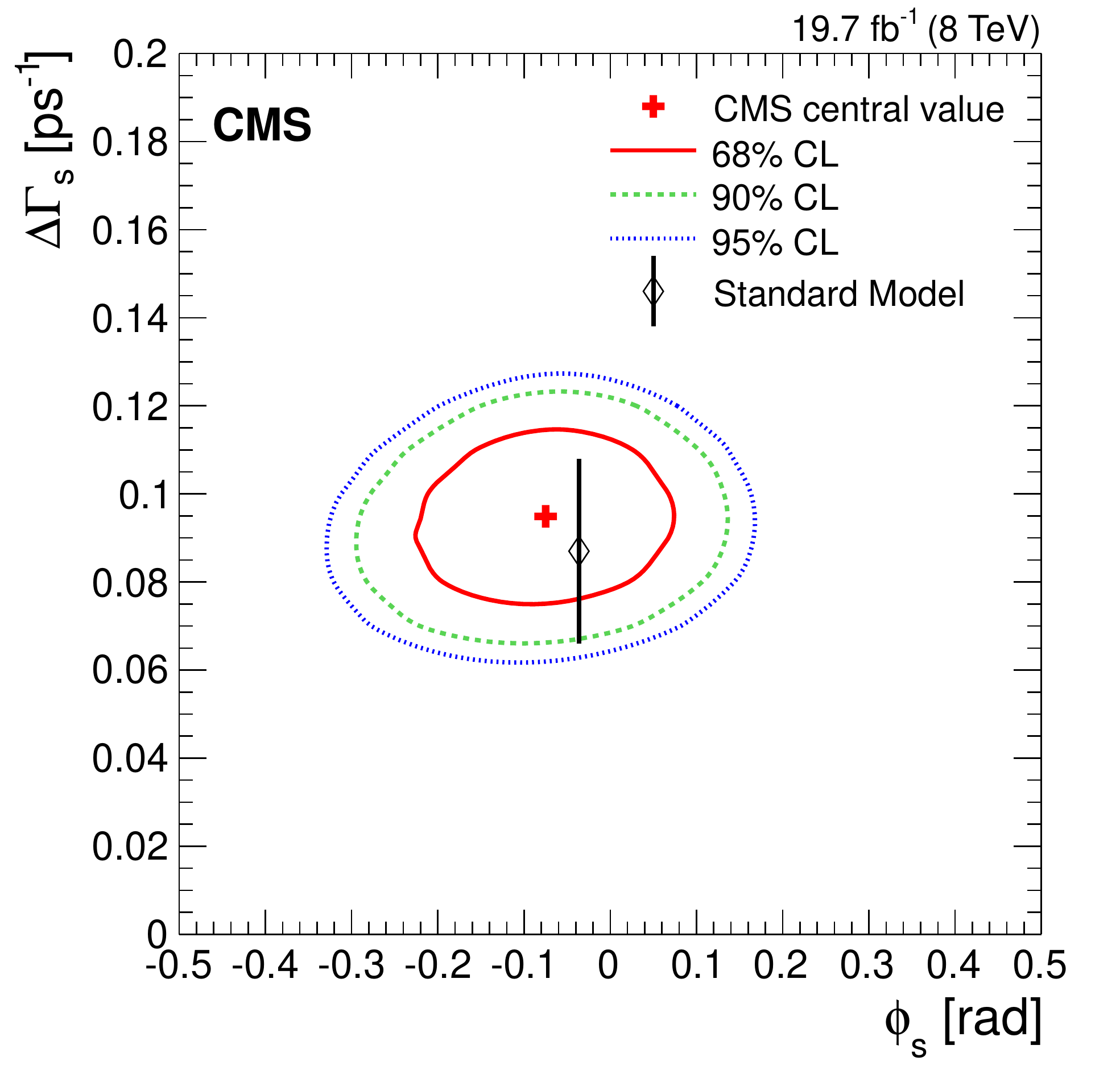}

\caption{The CMS measured central value and the 68\%, 90\%, and 95\% CL contours in the \deltagammas\ versus $\phi_\mathrm{s}$ plane, together with the SM prediction~\cite{Charles:2011va,Lenz:2011ti}. Uncertainties are statistical only.}
\label{fig:2dcontour}
\end{figure}

The results of the fit are given in Table~\ref{table:datatagged}, where the quoted uncertainties are statistical only. The corresponding correlation matrix for the statistical uncertainties in the physics fit parameters is shown in Table~\ref{table:datacorre}.
Since the likelihood profiles of $\delta_{\parallel}$, $\delta_{S\perp}$, and $\abs{A_S}^2$ are not parabolic, the statistical uncertainties quoted for these parameters are found from the increase in $-\log\cal{L}$ by 0.5.
In the fit, the value of $\Delta m_\mathrm{s}$ is allowed to vary following a Gaussian distribution with mean and standard deviation set to $(17.69\pm 0.08)\times 10^{12}~\hbar/\mathrm{s}$~\cite{Beringer:1900zz}.
As a cross-check, the $\Delta m_\mathrm{s}$ value is also left free to float and its best fit value is found to be in statistical agreement with the set value.
The various data distributions and the fit projections are shown in Figs.~\ref{fig:mass}, \ref{fig:lifetime}, and \ref{fig:datafittagged}. The drop in the $\cos\theta_\mathrm{T}$ distribution at the range limits is identified as being caused by close-by, high-angle kaon tracks. The central value and the 68\%, 90\%, and 95\% confidence level (CL) likelihood contours of the fit in the $\Delta\Gamma_\mathrm{s}$--$\phi_\mathrm{s}$ plane are shown in Fig.~\ref{fig:2dcontour}.

\begin {table}[h!t]
\centering
\topcaption{Results of the fit to the data. Uncertainties are statistical only.}
\label{table:datatagged}
  \begin{tabular}{ l | c }
    Parameter                                           & Fit result     \\
    \hline
    $\phi_\mathrm{s}~\mathrm{[rad]}$                    & $-0.075\pm 0.097$  \\
    $\Delta\Gamma_\mathrm{s}\,[\mathrm{ps}^{-1}]$        & $0.095\pm 0.013$  \\
    $\abs{A_0}^2$                                           & $0.510\pm 0.005$  \\
    $\abs{A_S}^2$                                           & $0.012\,_{-0.007}^{+0.009}$ \\
    $\abs{A_{\perp}}^2$                                     & $0.243\pm 0.008$  \\
    $\delta_{\parallel}~\mathrm{[rad]}$                 & $3.48\,_{-0.09}^{+0.07}$ \\
    $\delta_{S\perp}~\mathrm{[rad]}$                    &  $0.37\,_{-0.12}^{+0.28}$ \\
    $\delta_{\perp}~\mathrm{[rad]}$                     & $2.98\pm 0.36$  \\
    $c\tau\,[\mu\mathrm{m}]$                             & $447.2\pm 2.9$  \\
  \end{tabular}
\end{table}

\begin {table*}
\centering
\topcaption{Correlation matrix for the statistical uncertainties in the physics fit parameters.}
\label{table:datacorre}
  \begin{tabular}{ l | c  c c c c c c c c}
                                    & $\abs{A_0}^2$   & $\abs{A_S}^2$ &  $\abs{A_{\perp}}^2$  &  $\delta_{\parallel}$ & $\delta_{S\perp}$ & $\delta_{\perp}$  &   $c\tau$  & $\Delta\Gamma_\mathrm{s}$  & $\phi_\mathrm{s}$ \\
    \hline
    $|A_0|^2$                              		  & +1.00 & +0.19 & -0.64 & -0.08 & -0.18 & -0.02 & +0.38 & +0.70 & +0.11 \\
    $|A_S|^2$                               		 & \NA          & +1.00 & -0.02 & -0.32 & -0.79 & -0.10 & -0.16 & +0.01 & +0.03 \\
    $|A_{\perp}|^2$                         	 & \NA          &  \NA          & +1.00 & -0.27 & +0.03 & -0.06 & -0.50 & -0.77 & -0.11 \\
    $\delta_{\parallel}$              		 & \NA          &  \NA          & \NA          & +1.00 & +0.26 & +0.21 & +0.11 & +0.03 & -0.02 \\
    $\delta_{S\perp}$               		 & \NA          & \NA          & \NA          & \NA          & +1.00 & +0.06 & +0.11 & -0.04 & -0.06 \\
    $\delta_{\perp}$            		          & \NA          & \NA          & \NA          & \NA          & \NA          & +1.00 & +0.03 & +0.01 & +0.01 \\
    $c\tau$                                                & \NA          & \NA          & \NA          & \NA          & \NA          & \NA          & +1.00 & +0.55 & +0.10 \\
    $\Delta\Gamma_\mathrm{s}$          & \NA          & \NA          & \NA          & \NA          & \NA          & \NA          & \NA          & +1.00 & +0.10 \\
    $\phi_\mathrm{s}$                             & \NA          & \NA          & \NA          & \NA          & \NA          & \NA          & \NA          & \NA          & +1.00 \\
  \end{tabular}
\end{table*}

Several sources of systematic uncertainties in the primary measured quantities
are investigated by testing the various assumptions made in the fit model and those associated with the fit procedure.

The systematic uncertainty associated with the assumption of a constant efficiency as a function of $ct$ is evaluated by fitting the data with an alternative $ct$ efficiency parametrisation, which takes into account a small contribution of the decay time significance requirement at small $ct$ and first-order polynomial variations at high $ct$. The differences found in the fit results with respect to the nominal fit are used as systematic uncertainties.

The uncertainties associated with the variables $\cos\theta_\mathrm{T},\varphi_\mathrm{T}$, and $\cos\psi_\mathrm{T}$ of the 3D angular efficiency function are propagated to the fit results by varying the corresponding parameters within their statistical uncertainties, accounting for the correlations among the parameters.
The maximum variation of the parameters extracted from the fit is taken as the systematic uncertainty. The systematic uncertainty owing to a small discrepancy in the kaon \pt spectrum between data and simulation is evaluated by  weighting the events to make the simulated kaon \pt spectrum match that in data.

The uncertainty in the  $ct$ resolution associated with the $\kappa$ factor is propagated to the results.
A set of test samples is produced with the $\kappa(ct)$ factor varying within their uncertainty, assumed to be Gaussian.
One standard deviation of the distribution describing the difference between the $ct$ resolution with the nominal fit and with a varying $\kappa(ct)$ is taken as the systematic uncertainty.
Since the $\kappa(ct)$ factor is obtained from simulation, the associated systematic uncertainty is assessed by using a sample of prompt \JPsi{} decays obtained with an unbiased trigger and comparing them to similarly processed simulated data. In this way the decay time resolution for $ct\approx 0$ is obtained. The $\kappa(ct)$ factor is varied within the values observed in data and simulation. The resulting variations of the physics parameters are taken as systematic uncertainties.

Although the likelihood function makes use of a per-event mistag parameter, it does not contain a pdf model for the mistag distribution.
The associated systematic uncertainty is estimated by generating simulated pseudo-experiments with different mistag distributions for signal and background and fitting them with the nominal fit.

The dominant tagging systematic uncertainty originates from the assumption that the signal and calibration channels have the same tagging performance. It is evaluated using a calibration curve, obtained from simulated samples, that describes the mistag probability of \Bs{} as function of the mistag probability of \Bu. The fit to the data is repeated, re-calibrating the mistag probability with the \Bs--\Bu{} calibration curve, and the differences found in the fit results with respect to the nominal fit are used as the systematic uncertainties.

Possible biases intrinsic to the fit model are also taken into account. The nominal model function is tested by using simulated pseudo-experiments, and the average of the pulls (defined as the difference between the result of fit to the pseudo-experiment sample and the nominal value) is used as a systematic uncertainty if it exceeds one standard deviation statistical uncertainty.

The various hypotheses that have been assumed when building the likelihood function are tested by generating simulated pseudo-experiments with different hypotheses and fitting the samples with the nominal likelihood function. The obtained pull histograms of the physics variables are fitted with Gaussian functions, and the average of the pull is used as a systematic uncertainty if the difference with respect to the average exceeds one standard deviation statistical uncertainty. Concerning the modelling of the $\JPsi\,\PKp\PKm$ invariant mass distribution, the background model is changed to a Chebyshev function from the nominal exponential pdf. The  $ct$ background pdf is changed to the sum of three exponential functions instead of the two exponential functions of the nominal fit. The angular background pdf is generated by using the background simulation angular shapes instead of the fit ones. The effect of not including the angular resolution is also tested, using the residual distributions obtained from simulations. The RMS of the angular resolutions were found to be 5.9, 6.3, and 10 mrad, for $\cos\theta_T$, $\cos\psi_T$, and $\varphi_T$ respectively. The contribution to the systematic uncertainty from the background tagging asymmetry is negligible.

The hypothesis that $\abs{\lambda}=1$ is tested by leaving that parameter free in the fit.
The obtained value of $\abs{\lambda}$ is consistent with 1.0 within one standard deviation.
The differences found in the fit results with respect to the nominal fit are used as systematic uncertainties.

The alignment systematic uncertainty affects the vertex reconstruction and therefore the decay times. That effect is estimated to be 1.5\mum from studies of known B hadron lifetimes~\cite{Chatrchyan:2013sxa}.  The systematic effect owing to the very small number of \Bs{} originating from $\mathrm{B_c^+}\to\Bs{}\Pgpp$ feed-down, which would be reconstructed with large values of $ct$, has been found to be negligible.

The measured values for the weak phase $\phi_\mathrm{s}$ and the decay width difference \deltagammas{} are:
\begin{equation*}
\begin{split}
\phi_\mathrm{s} &= -0.075 \pm 0.097\stat\pm 0.031\syst\unit{rad},\\
\Delta\Gamma_\mathrm{s}&=  0.095 \pm 0.013\stat\pm 0.007\syst\unit{ps}^{-1}.
\end{split}
\end{equation*}
The systematic uncertainties are summarised in Table~\ref{table:systematics}. The uncertainties in the $\phi_\mathrm{s}$ and \deltagammas\ results are dominated by the statistical uncertainties.

\begin {table*}[h!tb]
\centering
\topcaption{Summary of the uncertainties in the measurements of the various \Bs{} parameters. If no value is reported, then the systematic uncertainty is negligible with respect to the statistical and other systematic uncertainties. The total systematic uncertainty is the quadratic sum of the listed systematic uncertainties.}
\label{table:systematics}
\resizebox{\textwidth}{!}{
  \begin{tabular}{@{\hskip 0pt} l@{\hskip 10pt} c@{\hskip 5pt}  c@{\hskip 5pt}  c@{\hskip 5pt}  c@{\hskip 5pt}  c@{\hskip 5pt}  c@{\hskip 5pt}  c@{\hskip 5pt}  c@{\hskip 5pt}  c @{\hskip 0pt}}
    Source of uncertainty           & $\phi_\mathrm{s}\,[\text{rad}]$   & $\Delta\Gamma_\mathrm{s}\,[\text{ps}^{-1}]$ & $\abs{A_0}^2$  &  $\abs{A_S}^2$  & $\abs{A_{\perp}}^2$   &  $\delta_{\parallel}\,[\text{rad}]$ &   $\delta_{S\perp}\,[\text{rad}]$ &   $\delta_{\perp}\,[\text{rad}]$  & $c\tau\,[\mu \mathrm{m}]$ \\
    \hline
    $ct$ efficiency                              & 0.002               & 0.0057                                         & 0.0015        &  \NA                   &  0.0023                 & \NA                                             & \NA                                             & \NA                                              & 1.0 \\
    Angular efficiency                                   & 0.016               & 0.0021                                         & 0.0060        &  0.008          &  0.0104                 & 0.674                                    & 0.14                                      & 0.66                                        & 0.8 \\
    Kaon \pt weighting    & 0.014                & 0.0015                                         & 0.0094        &  0.020         &  0.0041                 & 0.085                                     & 0.11                                     & 0.02                                         & 1.1 \\
    $ct$ resolution                              & 0.006               & 0.0021                                          & 0.0009        & \NA                   & 0.0008                  & 0.004                                    & \NA                                             & 0.02                                         & 2.9 \\
    Mistag distribution modelling          & 0.004               & 0.0003                                          & 0.0006        & \NA                    & \NA                             & 0.008                                     & 0.01                                       & \NA                                              & 0.1\\
    Flavour tagging                              & 0.003               & 0.0003                                          & \NA                    &  \NA                   & \NA                            & 0.006                                            & 0.02                                             & \NA                                 & \NA \\
    Model bias                                           &  0.015              & 0.0012                                          & 0.0008        & \NA                    & \NA                              & 0.025                                    & 0.03                                      & \NA                                                & 0.4 \\
    pdf modelling assumptions            & 0.006               &  0.0021                                          &  0.0016       &  0.002           &  0.0021                &  0.010                                    &  0.03                                   &  0.04                                         &     0.2\\
    $\abs{\lambda}$ as a free parameter   & 0.015               & 0.0003                                           & 0.0001        &  0.005         &  0.0001                   & 0.002                                     & 0.01                                     & 0.03                                         & \NA \\
    Tracker alignment  & \NA               & \NA                                           & \NA        &  \NA         &  \NA                   & \NA                                     & \NA                                     & \NA                                         & 1.5 \\
    \hline
    Total systematic uncertainty          & 0.031               & 0.0070                                          & 0.0114         & 0.022          & 0.0116                    &  0.680                                  &    0.18                            &   0.66                                        & 3.7  \\ \hline
    Statistical uncertainty                      & 0.097               & 0.0134                                           & 0.0053        & 0.008           & 0.0075                   & 0.081                                    & 0.17                                      & 0.36                                           & 2.9\\
  \end{tabular}
}
\end{table*}

\section{Summary}
Using pp collision data collected by the CMS experiment at a centre-of-mass energy of  8\TeV and corresponding to  an integrated luminosity of 19.7\fbinv, 49\,200 \BsJpsiPhi{} signal candidates were used to measure the weak phase $\phi_\mathrm{s}$ and the decay width difference \deltagammas. The analysis was performed by using opposite-side lepton tagging of the \Bs{} flavour at the production time. Both muon and electron tags were used.

The measured values for the weak phase and the decay width difference between the \Bs{} mass eigenstates are $\phi_\mathrm{s} = -0.075 \pm 0.097\stat\pm 0.031\syst\unit{rad}$ and $\Delta\Gamma_\mathrm{s}=  0.095 \pm 0.013\stat\pm 0.007\syst\unit{ps}^{-1}$, respectively. The measured values are consistent with those obtained by the LHCb Collaboration using \Bs{}$\to$J$/\psi\PKp\PKm$ decays~\cite{PhysRevLett.114.041801}.

Our measured value of $\phi_\mathrm{s}$ agrees with the SM prediction.
Our result confirms \deltagammas\ to be nonzero, with a value consistent with theoretical predictions.
The uncertainties in our $\phi_\mathrm{s}$ and \deltagammas\ measurements are dominated by statistical uncertainties. Our results provide independent reference measurements of $\phi_\mathrm{s}$ and \deltagammas,
and contribute to improving the overall precision of these quantities and thereby probing the SM
further. Since our measurement precision is still limited by statistical uncertainty, substantial improvement is expected from LHC $\sqrt{s}=13$\TeV high-luminosity running that will be available over the next few years.

\begin{acknowledgments}
We congratulate our colleagues in the CERN accelerator departments for the excellent performance of the LHC and thank the technical and administrative staffs at CERN and at other CMS institutes for their contributions to the success of the CMS effort. In addition, we gratefully acknowledge the computing centres and personnel of the Worldwide LHC Computing Grid for delivering so effectively the computing infrastructure essential to our analyses. Finally, we acknowledge the enduring support for the construction and operation of the LHC and the CMS detector provided by the following funding agencies: BMWFW and FWF (Austria); FNRS and FWO (Belgium); CNPq, CAPES, FAPERJ, and FAPESP (Brazil); MES (Bulgaria); CERN; CAS, MOST, and NSFC (China); COLCIENCIAS (Colombia); MSES and CSF (Croatia); RPF (Cyprus); MoER, ERC IUT and ERDF (Estonia); Academy of Finland, MEC, and HIP (Finland); CEA and CNRS/IN2P3 (France); BMBF, DFG, and HGF (Germany); GSRT (Greece); OTKA and NIH (Hungary); DAE and DST (India); IPM (Iran); SFI (Ireland); INFN (Italy); MSIP and NRF (Republic of Korea); LAS (Lithuania); MOE and UM (Malaysia); CINVESTAV, CONACYT, SEP, and UASLP-FAI (Mexico); MBIE (New Zealand); PAEC (Pakistan); MSHE and NSC (Poland); FCT (Portugal); JINR (Dubna); MON, RosAtom, RAS and RFBR (Russia); MESTD (Serbia); SEIDI and CPAN (Spain); Swiss Funding Agencies (Switzerland); MST (Taipei); ThEPCenter, IPST, STAR and NSTDA (Thailand); TUBITAK and TAEK (Turkey); NASU and SFFR (Ukraine); STFC (United Kingdom); DOE and NSF (USA).

Individuals have received support from the Marie-Curie programme and the European Research Council and EPLANET (European Union); the Leventis Foundation; the Alfred P. Sloan Foundation; the Alexander von Humboldt Foundation; the Belgian Federal Science Policy Office; the Fonds pour la Formation \`a la Recherche dans l'Industrie et dans l'Agriculture (FRIA-Belgium); the Agentschap voor Innovatie door Wetenschap en Technologie (IWT-Belgium); the Ministry of Education, Youth and Sports (MEYS) of the Czech Republic; the Council of Science and Industrial Research, India; the HOMING PLUS programme of the Foundation for Polish Science, cofinanced from European Union, Regional Development Fund; the Compagnia di San Paolo (Torino); the Consorzio per la Fisica (Trieste); MIUR project 20108T4XTM (Italy); the Thalis and Aristeia programmes cofinanced by EU-ESF and the Greek NSRF; the National Priorities Research Program by Qatar National Research Fund; the Rachadapisek Sompot Fund for Postdoctoral Fellowship, Chulalongkorn University (Thailand); and the Welch Foundation.
\end{acknowledgments}

\bibliography{auto_generated}

\cleardoublepage \appendix\section{The CMS Collaboration \label{app:collab}}\begin{sloppypar}\hyphenpenalty=5000\widowpenalty=500\clubpenalty=5000\textbf{Yerevan Physics Institute,  Yerevan,  Armenia}\\*[0pt]
V.~Khachatryan, A.M.~Sirunyan, A.~Tumasyan
\vskip\cmsinstskip
\textbf{Institut f\"{u}r Hochenergiephysik der OeAW,  Wien,  Austria}\\*[0pt]
W.~Adam, E.~Asilar, T.~Bergauer, J.~Brandstetter, E.~Brondolin, M.~Dragicevic, J.~Er\"{o}, M.~Flechl, M.~Friedl, R.~Fr\"{u}hwirth\cmsAuthorMark{1}, V.M.~Ghete, C.~Hartl, N.~H\"{o}rmann, J.~Hrubec, M.~Jeitler\cmsAuthorMark{1}, V.~Kn\"{u}nz, A.~K\"{o}nig, M.~Krammer\cmsAuthorMark{1}, I.~Kr\"{a}tschmer, D.~Liko, T.~Matsushita, I.~Mikulec, D.~Rabady\cmsAuthorMark{2}, B.~Rahbaran, H.~Rohringer, J.~Schieck\cmsAuthorMark{1}, R.~Sch\"{o}fbeck, J.~Strauss, W.~Treberer-Treberspurg, W.~Waltenberger, C.-E.~Wulz\cmsAuthorMark{1}
% ...
\vskip\cmsinstskip
% ...
1:~~Also at Vienna University of Technology, Vienna, Austria\\
2:~~Also at CERN, European Organization for Nuclear Research, Geneva, Switzerland\\
% ...

\end{sloppypar}
\end{document}